\documentclass[aps,prl,twocolumn,groupedaddress]{revtex4}
\usepackage{graphicx}
\usepackage{amsmath,amsfonts,amssymb}
\usepackage{color}
\usepackage{float}
\begin{document}
\title{Tetrahedron of flavors: \\
One Three to rule them all}
\author{Aharon Davidson}
\email{davidson@bgu.ac.il}
\homepage{https://physics.bgu.ac.il/~davidson}
\affiliation{Physics Department, Ben-Gurion University
of the Negev, Beer-Sheva 84105, Israel}

\date{October10, 2022}

\begin{abstract}
	Under the conception that the total number
	three of fermion families must have the one
	and the same gauge theoretical origin as all
	other threes which accompany the single family
	grand unifiable group structure, we trade the
	trinification $SU(3)$ symmetry building block
	for its semi-simple Vertical-Horizontal
	symmetric $SU(3)_V \otimes SU(3)_H$
	extension.
	The anomaly free flavor-chiral fermion
	representation is then constructed, solely out of
	threes and without any superfluous replication,
	by treating each standard $V$-triplet (and anti
	triplet) as an $H$-singlet, and conversely, by
	letting each standard $V$-singlet transform
	as an $H$-triplet (or anti triplet).
	The model can be schematically described by
	a tetrahedron of flavors.
	In its bi-trinification phase, the model exhibits
	two coupled trinification cycles, sharing the same
	color group, and is furthermore accompanied by
	a built-in dark sector.
	In its isomorphic quartification phase, the model
	presents a novel non-Pati-Salam version of
	quark/lepton correspondence, where quarks
	are paired with anti-leptons to cancel horizontal
	anomalies, and all fermion masses stem from one
	and the same Yukawa source.
\end{abstract}

%\pacs{}

\maketitle

\section{Introduction}
Decoding the mass spectrum of quarks and
leptons is without any doubt the holy grail of
theoretical particle physics.
While the mass generating Higgs mechanism,
mediated by Yukawa couplings, is rooted in the
heart of the standard electro/nuclear theory,
we still do not have the slightest idea what
physics actually determines the eigenmasses
and mixings. 
At this stage, even an empirical formula to
account for the observed mass hierarchy is
certainly welcome.
The structure of the Fermi mass matrix is
however just the tip of the flavor puzzle
iceberg.

The various fermionic pieces which make a single family
(or a generation) are well organized, up to some
electro/nuclear neutral members, within the framework of
the by now standard
$SU(3)_C \otimes SU(2)_L \otimes U(1)_Y$ gauge theory.
Anomaly cancellation is then a necessary self consistency
condition, but obviously not a sufficient one for shedding
light on the full picture.
The consequent grand unification procedure
to combine the otherwise separated fermionic pieces
into a single representation has been exclusively realized
for the $SU(5) \subset SO(10)\subset E(6)$ group sequence,
and is most welcome, but leaves a fundamental question
unanswered.
Namely, why can/must grand unification be realized
(at least partially) already at the single family level,
while apparently leaving the observed multiplicity of
fermion families field/group theoretically out of the
game.

At the time, embedding the single family GUT within
a larger multi family GUT seemed to be a logical step
forwards.
Unfortunately, such a potentially promising direction
has been practically exhausted, and so far failed to deliver.
On the short list of theoretical (on top of experimental)
obstacles encountered one can find:

\noindent(i) The magic yet theoretically challenging
number of exactly \emph{three} standard fermion
families.

\noindent (ii) The restrictive chiral flavor structure
required at the low energy group theoretical reduction
level.

\noindent(iii) The ever growing sizes (counting
generators) of the candidate GUT groups involved.

\noindent(iv) Heavy field theoretical artillery, such
as super-symmetry and dimensional reduction, 
while being attractive from various other reasons,
has not contributed so far its part to the flavor puzzle.
The more so superstring theory, including its string
phenomenology outer branch, has not been proven
too useful either.

It may well be that
the tempting strategy of passing through the single
family grand unification stage may have counter
intuitively blocked us from unveiling the full flavor
picture.

\section{
Vertical $\leftrightarrow$ Horizontal symmetry}

\begin{figure}[h]
	\centering \bigskip
	\includegraphics[scale=0.18]{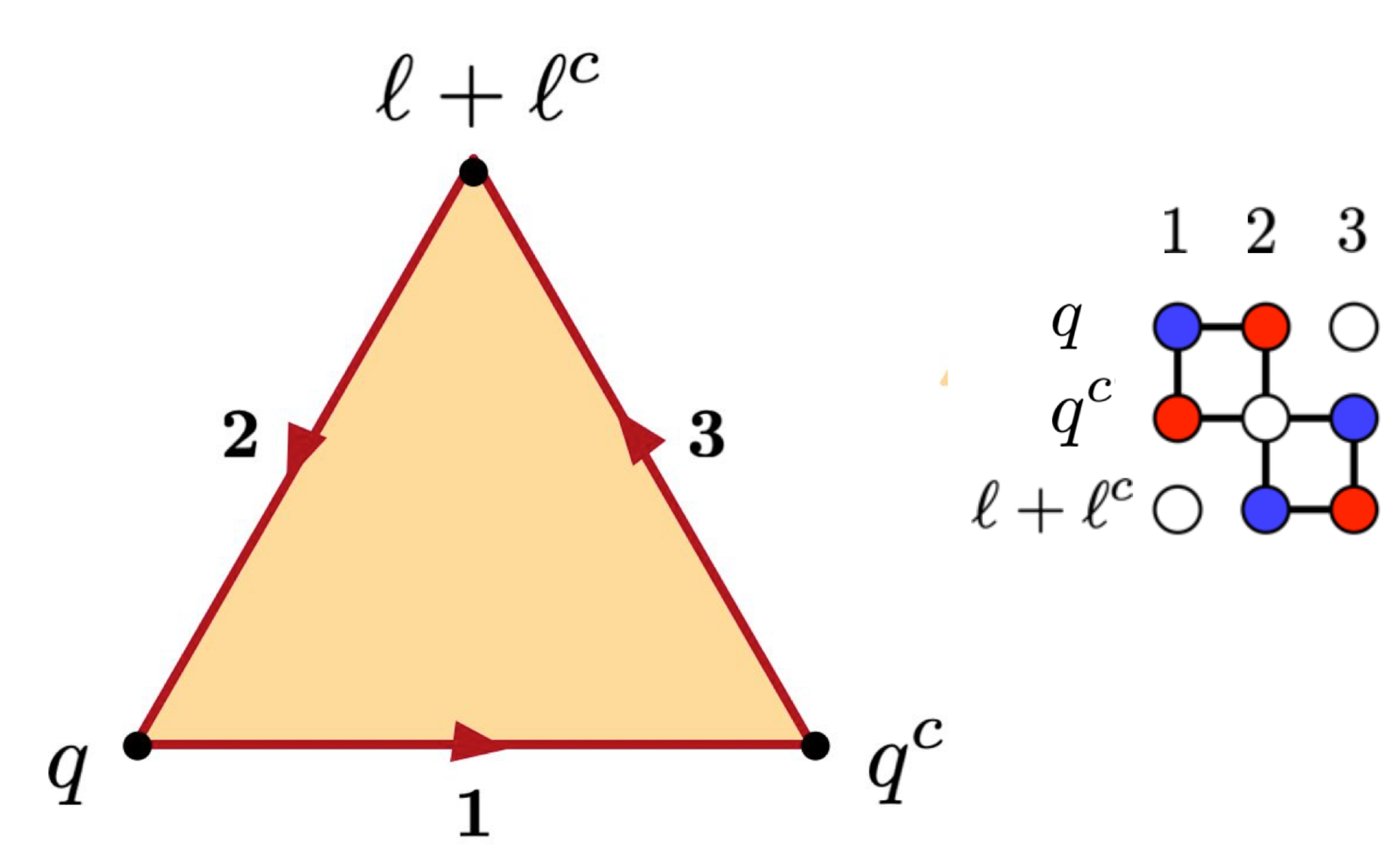}
	\caption{The trinification model:
	The vertices are associated with the various
	fermions involved., and the three edges stand
	for the three $SU(3)$ group factors involved.
	An incoming/outgoing arrow signals a
	triplet/anti-triplet of the respective $SU(3)$.
	Schematically, blue and red circles represent
	triplets and anti-triplets respectively}
	\label{tri} 
\end{figure}

Equipped with no compelling answers, the flavor
puzzle has sourced plenty of imaginative ideas,
ranging from atomic style isotopes all the way
to compositeness, involving a variety of field
theoretical techniques.
In this paper, however, with the focus solely on
group theory, we humbly follow the hypothesis
that the total number \emph{three} of fermion
families shares the one and the same gauge
theoretical origin as all other \emph{threes}
which govern the single family unifiable (not
grand unified) group structure.
Adopting this line of thought, let our starting
point be trinification \cite{333,trinification},
based on the semi-simple gauge group
\begin{equation}
	G=SU(3)_C \otimes SU(3)_L 
	\otimes SU(3)_R ~.
\end{equation}
An accompanying $C_3$ discrete symmetry is
optional, and becomes mandatory if a common
gauge coupling constant is in order.
Associated with is the familiar anomaly free flavor
chiral representation of left handed fermions
\begin{equation}
	\psi_L=
	\begin{array}{|c ||c|c|c|}
	\hline q & 3 & 3^\ast & 1 \\
	\hline q^c & 3^\ast & 1 & 3 \\
	\hline \ell+\ell^c & 1 & 3 & 3^\ast \\
	\hline \end{array}
	\label{333}
\end{equation}
which can always be supplemented, if so required,
by a bunch of dark $G$-singlets
\begin{equation}
	\begin{array}{|c ||c|c|c|}
	\hline ~~\chi~~ & ~1~ & ~1~ & ~1~ \\
	\hline \end{array} 
\end{equation}
Schematically, the model can be neatly represented
by a triangle, as depicted in Fig.\ref{tri}, where the
vertices represent fermions and the edges stand
for the various $SU(3)$ group factors involved. 
%% Fig 1  %%%

Already at the trinification level, prior to $E(6)$
unification \cite{E6}, there is a heavy price
(not everyone is willing) to pay, namely extending
the standard single fermion family, composed of 16
members, to include extra 27-16=11 non-standard
members.
The latter supplement, easily classified by means
of the electric charge operator
\begin{equation}
	Q=T_{3L}+T_{3R}+\frac{1}{2}(Y_L+Y_R) ~,
	\label{Q}
\end{equation}
decouples however once $SU(3)_L \otimes SU(3)_R$
parent symmetry breaks down to its left-right symmetric
sub-group $SU(2)_L \otimes SU(2)_R \otimes U(1)_{B-L}$.
Furthermore, notice that the full quark/lepton
correspondence, i.e. treating lepton number as
the forth color, \cite{PS} is not restored until
(and if) $E(6)$ embedding is realized.

Multiplying the trinification group $G$ by extra
$SU(3)$ factors is then quite a conservative step
forwards.
$SU(3)^N$ models, with $N>3$, mostly single
family models, have been extensively studied by Ma
\cite{SUnk} and by others \cite{SU34}.
Horizontal symmetries \cite{horizontal}, primarily invoked to classify
the otherwise electro/nuclear degenerate fermionic
families, start from minimal $U(1)$, local or else global
(incorporating the Peccei-Quinn mechanism \cite{familon}) and in
particular include an $SU(3)$ example \cite{HSU3}.
Here, we revive the idea of  Vertical
$\leftrightarrow$ Horizontal symmetry \cite{VH}, and trade
the single family group, now referred to as $G_V$,
for its $G_V \otimes G_H$ semi-simple extension.
Note that models incorporating semi-simple group
structure, for example Pati-Salam's $SU(4)^4$ \cite{PS},
first ever unification scheme (originally designed for
two families), the hybrid single family left-right symmetric
$SU(5)_L \otimes SU(5)_R$ model \cite{5x5} hosting
chiral color \cite{chiralcolor}, and multi family models
of the $SO(10)\otimes SO(10)$ type \cite{10x10}, have
already been discussed in the literature.

To construct the tenable $G_V \otimes G_H$
fermion representation, we adopt the following
prescription:

\noindent \emph{Treat each standard $V$-triplet (and anti-triplet)
as an $H$-singlet, and conversely, let each standard
$V$-singlet transform as an $H$-triplet (or anti-triplet).}

\noindent This prescription generalizes Eq.(\ref{333}) into
\begin{equation}
\begin{array}{ccc}
	&~~\quad \quad  V \quad\quad\quad\quad H & \\
	& \psi_L=
	\begin{array}{|c |c |c ||c |c |c |}
	\hline 3 & 3^\ast & 1 & 1 & 1 & x \\
	\hline 3^\ast & 1 & 3 & 1 & y & 1 \\
	\hline 1& 3 & 3^\ast & z & 1 & 1 \\
	\hline\hline 1 & 1 & 1 & z^\ast &
	y^\ast & x^\ast \\
	\hline \end{array}  &   
\end{array}
	\label{rep} 
\end{equation}
The cancelation of ABJ anomalies is done
pairwise, while maintaining overall flavor chirality,
so that each of the $SU(3)$ representations
$x,y,z$ can be either a $3$ or else a $3^\ast$.
While the particular choice seems irrelevant
from any individual $SU(3)$ point of view,
a matter of definition, it does become relevant
in the presence of an accompanying discrete symmetry.
Altogether, with only $SU(3)$ triplets and
anti-triplets at our disposal, only two independent
configurations exist.

\begin{figure}[h]
	\centering \bigskip
	\includegraphics[scale=0.22]{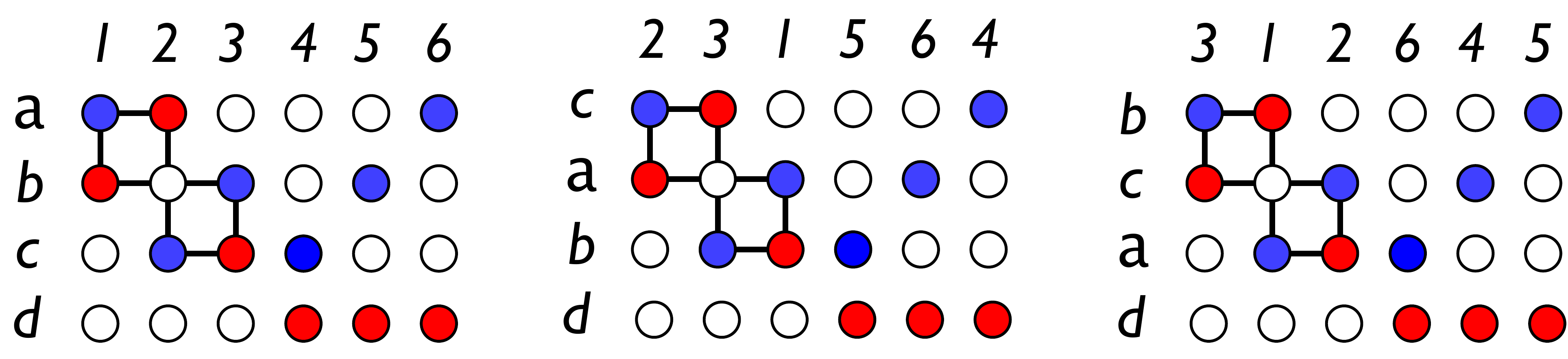}
	\caption{With blue and red circles
	representing triplets and anti-triplets
	respectively, the three trinification ways
	to identify $\{q,q^c,\ell+\ell^c\}$ are
	illustrated by $\{a,b,c\}$ cyclic permutations
	accompanied by suitable re-arrangements
	of the six $SU(3)$ group factors.}
	\label{3variants} 
\end{figure}

The manifestly symmetric configuration
\begin{equation}
	x=y=z \quad \textrm{} ~,
	\label{xyz1}
\end{equation} 
could have been our naive preference.
But unfortunately, as demonstrated in
Fig.(\ref{3variants}), its associated discrete
symmetry does not really go much beyond the
original trinity model, offering the same
three ways to identify $\{q,q^c,\ell+\ell^c\}$,
and this without involving $V\leftrightarrow H$
interplay, and without reviving quark/lepton
correspondence.

%% Fig 2  %%%

\section{Tetrahedron of flavors}

\begin{figure}[h]
	\centering \bigskip
	\includegraphics[scale=0.22]{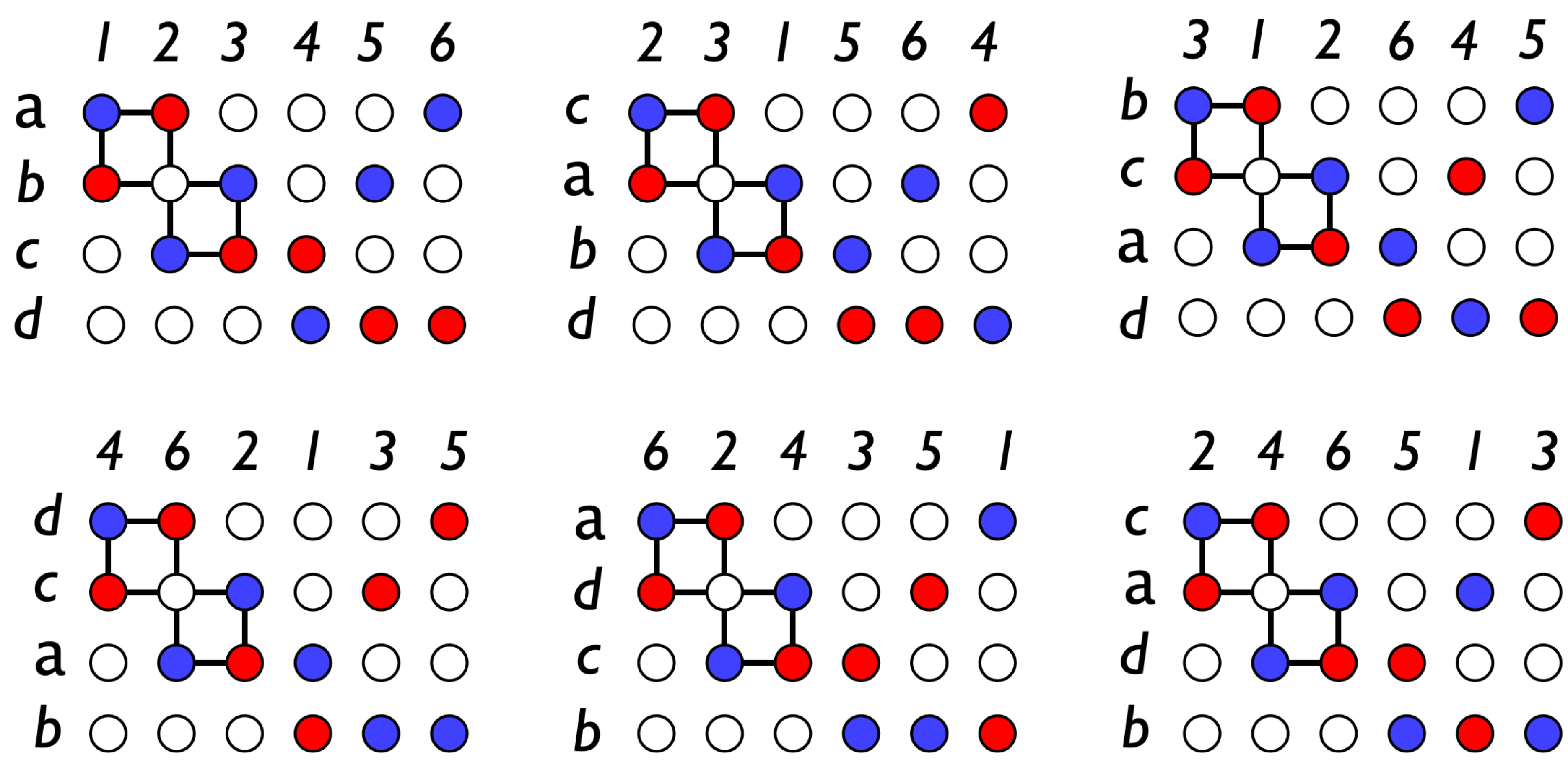}
	\caption{Horizontal bi-trinification:
	By rearranging raws (fermion assignments)
	and columns ($SU(3)$) group factors),
	while recasting the exact structure of
	the trinification V-sector (marked with lines),
	one can easily switch from one H-sector
	(three right columns) to another.
	The six variants are thus equivalent to each
	other.
	Note that triplets and anti-triplets are
	represented here by blue and red circles,
	respectively.}
	\label{6variants} 
\end{figure}

In this paper, however, from reasons to
be specified, we construct the
so-called bi-trinification model (along with
its isomorphic quartification phase) based
on the alternative configuration
\begin{equation}
	x=y= z^\ast  \quad
	\textrm{up to permutations} ~. 
	\label{xyz2}
\end{equation}
It offers six variant ways to identify
$\{q,q^c,\ell+\ell^c\}$, each of which comes
with its own (practically equivalent) permuted
and/or conjugate horizontal assignments.
To fully appreciate this point, as demonstrated
in Fig.\ref{6variants}, one may rearrange raws as
well as columns, identify an alternative trinification
V-sector (three left columns), and then verify how
one H-sector (three right columns) gets systematically
replaced by an equivalent one, such that
$(x, y, z) \rightarrow
(x^\prime, y^\prime, z^\prime)$, where the latter
is a permuted version of the former, with
or without conjugation.

%% Fig 3  %%%

%% Fig 4  %%%
\begin{figure}[h]
	\includegraphics[scale=0.26]{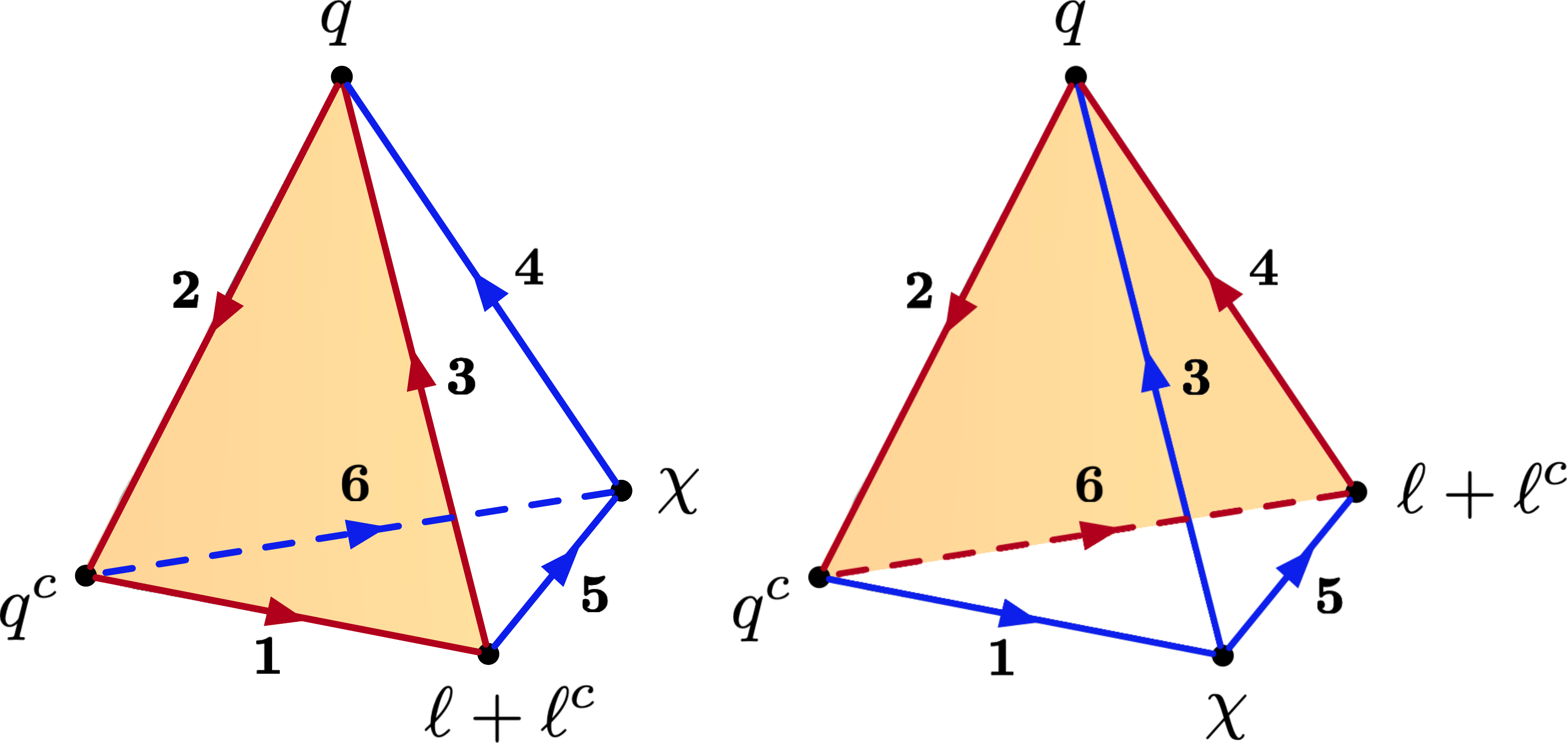}
	\caption{The Tetrahedron model
	(Bi-Trinification phases):
	The two coupled 3-cycles (red arrows), namely
	$231$ (left plot) and $246$ (right plot), describe
	two different trinification V-sectors.
	Attached to each one of them is a corresponding
	H-sector (blue arrows).
	In this phase,
	$\ell$ and $\ell^c$ conventionally share a common
	leptonic vertex, and associated with the fourth
	vertex $\chi$ is a novel dark (= electro/nuclear
	neutral) sector.}
	\label{3cycles} 
\end{figure}

The fermion representation furnishes a tetrahedron.
While the edges, numbered $k=1,2,...,6$, correspond
to the individual $SU(3)_k$ group factors, the vertices
are associated with the four fermion classes
$q, q^c,\ell+\ell^c,\chi$.
The various arrows involved, making the edges
directed, stand for triplets (outgoing arrows) and
anti-triplets (ingoing arrows).
Notice that there are in fact two
different ways to consistently identify the V-sector.
Sticking to Fig.(\ref{6variants}), it is either the
231-cycle (three upper configurations) or alternatively
the 246-cycle (three lower configurations).
It is this characteristic feature which justifies
using the terminology bi-trinification.
The two cycles correspond to the trinifications
$SU(3)_C \otimes \left[SU(3)_L\otimes SU(3)_R
\right]_{V,H}$, respectively.
This in turn gives rise to two electric charges
\begin{equation}
	Q_{V,H}=\left[ T_{3L}+T_{3R}
	+\frac{1}{2}Y_{L+R} \right]_{V,H} ~,
	\label{QVH}
\end{equation}
one of which, that is $Q_H$, must be spontaneously
broken, thereby opening the door for the Holdom
effect \cite{Holdom}, also known as photon mixing.

\begin{figure}[h]
	\includegraphics[scale=0.24]{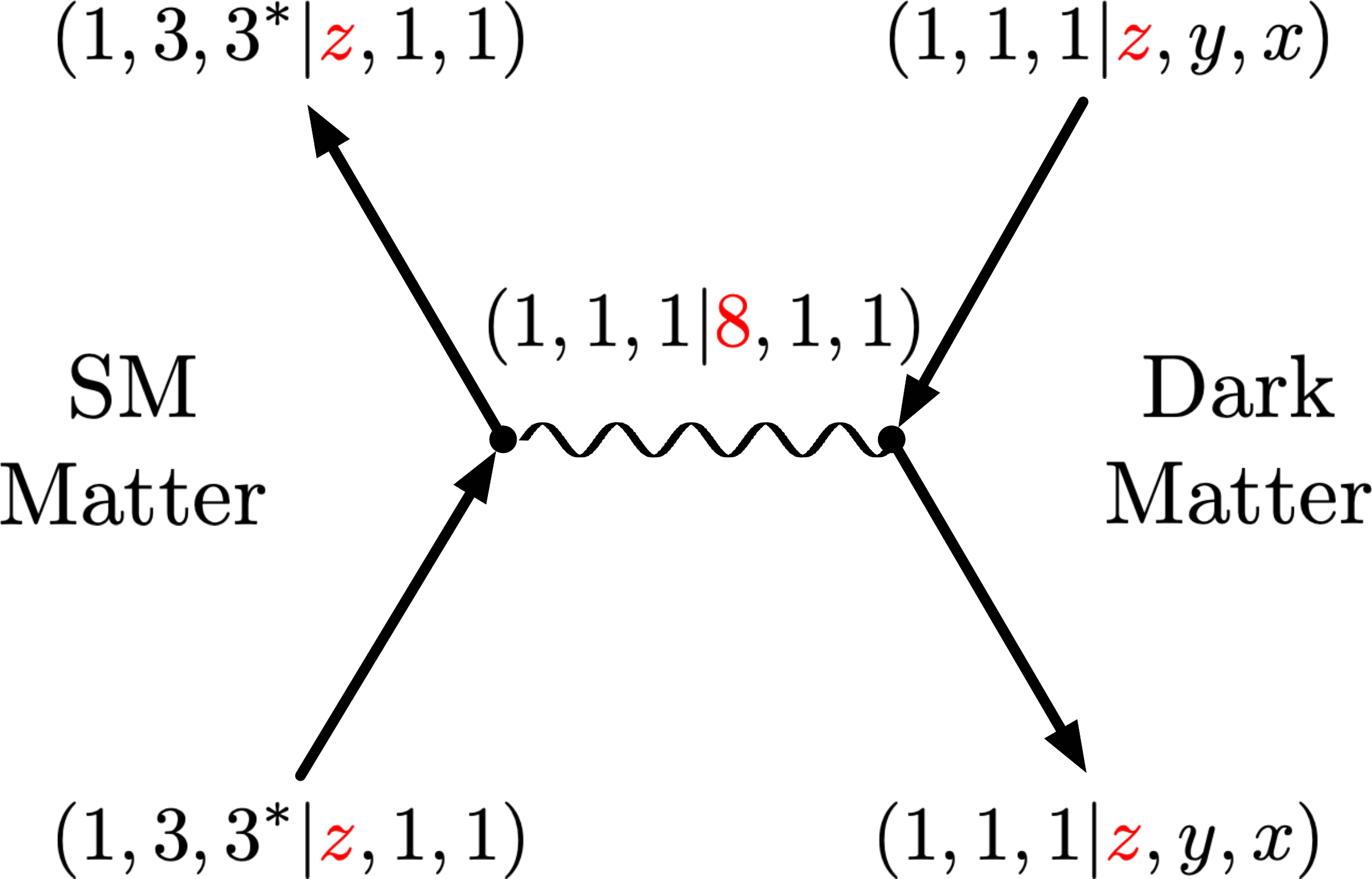}
	\caption{Dark matter portal:
	Dark matter particles ($V$-sector singlets)
	interact with standard model particles
	($V$-sector non singlets) by exchanging
	heavy horizontal gauge bosons.
	Depicted in this figure is the lepto/dark case.}
	\label{Portal} 
\end{figure}

The $\chi$-fermions are required on anomaly
cancelation grounds.
They are $V$-sector singlets by construction, and
as such, have no standard electro/nuclear interactions.
This makes them, by definition, candidates
for dark matter particles of the WIMP kind.
They do interact with ordinary matter though,
with the $H$-sector serving as the tenable portal.
The mechanism involved, as demonstrated in
Fig.(\ref{Portal}) for the special lepto/dark case,
is the exchange of super heavy horizontal gauge
bosons.

%% Fig 5  %%%

\section{Bi-Tri/Quart isomorphism}

\begin{figure}[h]
	\includegraphics[scale=0.34]{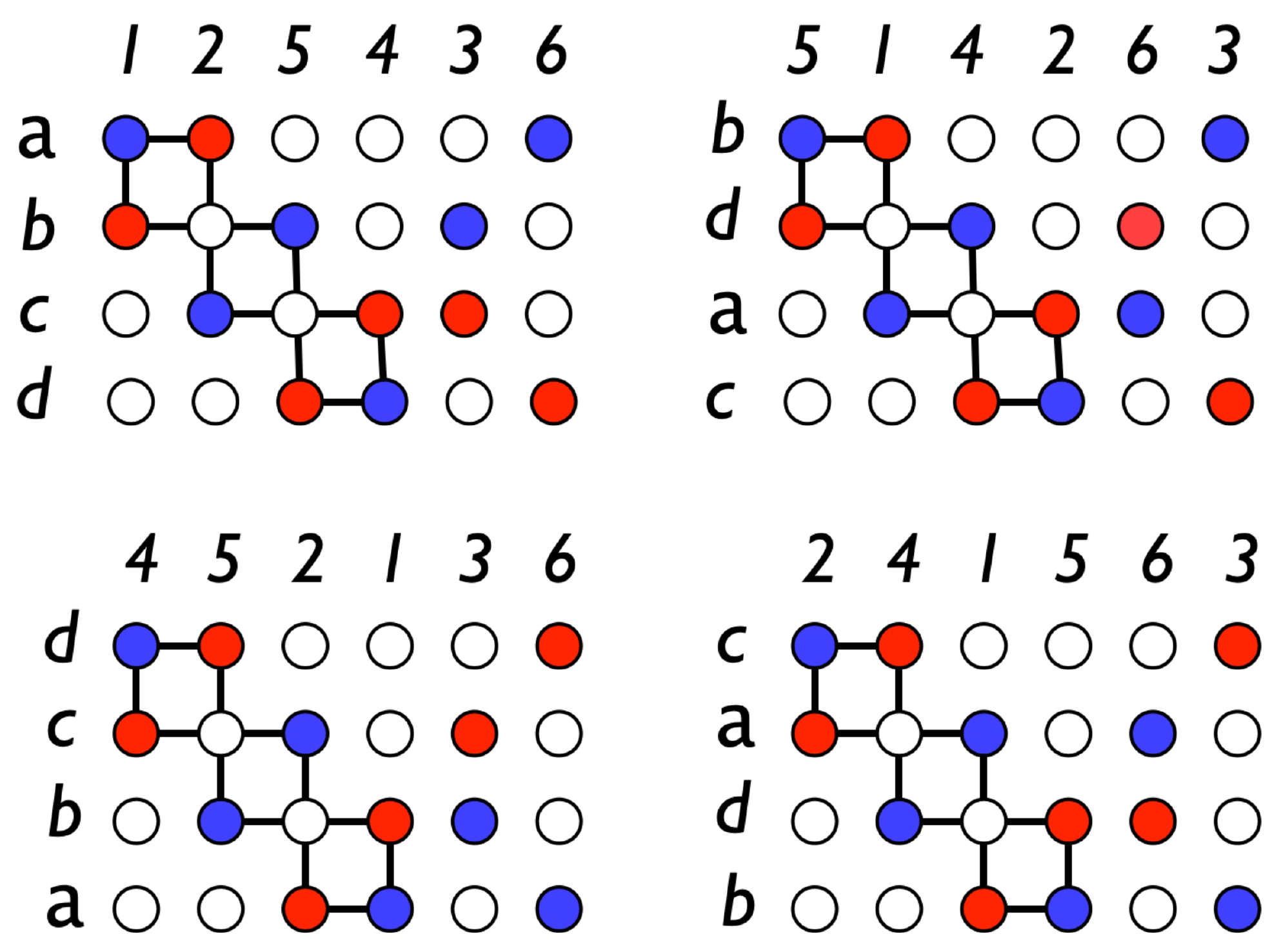}
	\caption{Horizontal quartification:
	Alternative rearrangements of particle
	assignments (raws) and $SU(3)$ group
	factors (columns) depart from trinification,
	and extend the V-sectior (marked with lines)
	into quartification at the expense of a smaller
	(two right columns) H-sector.
	The equivalence of these four variants to
	each other is interpreted as a novel version
	of quark/lepton correspondence.
	As before, triplets and anti-triplets are
	represented by blue and red circles, respectively.}
	\label{4variants} 
\end{figure}

Alternative rearrangements of raws and columns
can make the single family group flavor assignments
depart from their built-in trinification (and thus
bi-trinification) construction, and at the expense
of a smaller H-sector, extend the V-sectior into
a quartification phase.
Depicted in Fig.(\ref{4variants}) are the four
associated quartification variant configurations.
Indeed, they share a common extended  V-sector
(left four columns connected with lines) and admit
shrank (only two, rather than three, right columns)
horizontal structures. 
They are equivalent to each other.
And most importantly, the existence of these four
variants serves to establish that, as far as our model
is concerned, bi-trinification and quartification
are two faces of the same coin.

%% Fig 6  %%%

Back to the tetrahedron of flavors, the quartification
phase calls for some reshuffling of fermion assignments.
To be specific, the leptonic sub-representation bifurcates,
such that $\ell$ and $\ell^c$ belong now to two different
vertices.
To classify the fermions under the extended single
family group $SU(3)^4$, with the individual subscripts
being $C,L,R,N$ (with $N$ denoting the new comer),
we introduce the fully symmetric electric charge
formula
\begin{equation}
	Q=T_{3L}+T_{3R}+T_{3N}
	+\frac{1}{2}\left(Y_L+Y_R+Y_N \right) ~.
\end{equation}
Apart from the manifest $LRN$ symmetry, it has
the advantage that all individual electric charges
involved are standard.
The electric charge $Q$, which generalizes
Eq.(\ref{Q}), should be contrasted with
Eq.(\ref{QVH}), and with a different electric
charge formula recently advocated for a similar
$SU(3)^4$ model.
To be more explicit, and expose the extra nine
leptons per family, introduced on top of the
trinification scheme, we piecewise specify the
$SU(3)_C \otimes SU(2)_L \otimes SU(2)_R
\otimes U(1)_{B-L}$ content of the
fermion representation, namely
\begin{equation}
\begin{array}{ccl}
	(3,3^\ast,1,1)&  = &
	(3,2,1)_{\frac{1}{3}}
	+(3,1,1)_{-\frac{2}{3}} ~, \\
	(3^\ast,1,3,1)&  = &
	(3^\ast,1,2)_{-\frac{1}{3}}
	+(3^\ast,1,1)_{\frac{2}{3}} ~, \\
	(1,3,1,3^\ast)& =  &
	(1,2,1)_{-1,1} +(1,2,1)_{-1} ~+    \\
	& + & (1,1,1)_{0,2} +(1,1,1)_0  ~,   \\
	(1,1,3^\ast,3)& =  &
	(1,1,2)_{-1,1} + (1,1,2)_1~+ \\
	& + & (1,1,1)_{-2,0} +(1,1,1)_0  ~.
\end{array}
\end{equation}
Of special interest are those vector-like fermions
which carry non-standard $B-L$ charges.
From a group theory point of view, they are primarily
responsible for the small value
\begin{equation}
	\sin^2 \theta_W \rightarrow
	\frac{\sum T_{L3}^2}{\sum Q^2}=\frac{1}{4}
\end{equation}
of the Weinberg angle at the symmetry limit, which is
smaller than the conventional value of $\frac{3}{8}$.
However, the vector-like surplus of quarks and leptons
is not protected by the
$SU(3)_C \otimes SU(2)_L \otimes SU(2)_R
\otimes U(1)_{B-L}$ symmetry subgroup, and
decouples at some presumably very
heavy mass scale.
Under the latter residual symmetry, we are
left with the exact flavor chiral family of the
Left-Right symmetric model, that is
\begin{equation}
	\psi= (3,2,1)_{\frac{1}{3}}
	+(3^\ast,1,2)_{-\frac{1}{3}}+
	(1,2,1)_{-1}+ (1,1,2)_1 ~.
\end{equation}

%% Fig 6  %%%
\begin{figure}[h]
	\includegraphics[scale=0.2]{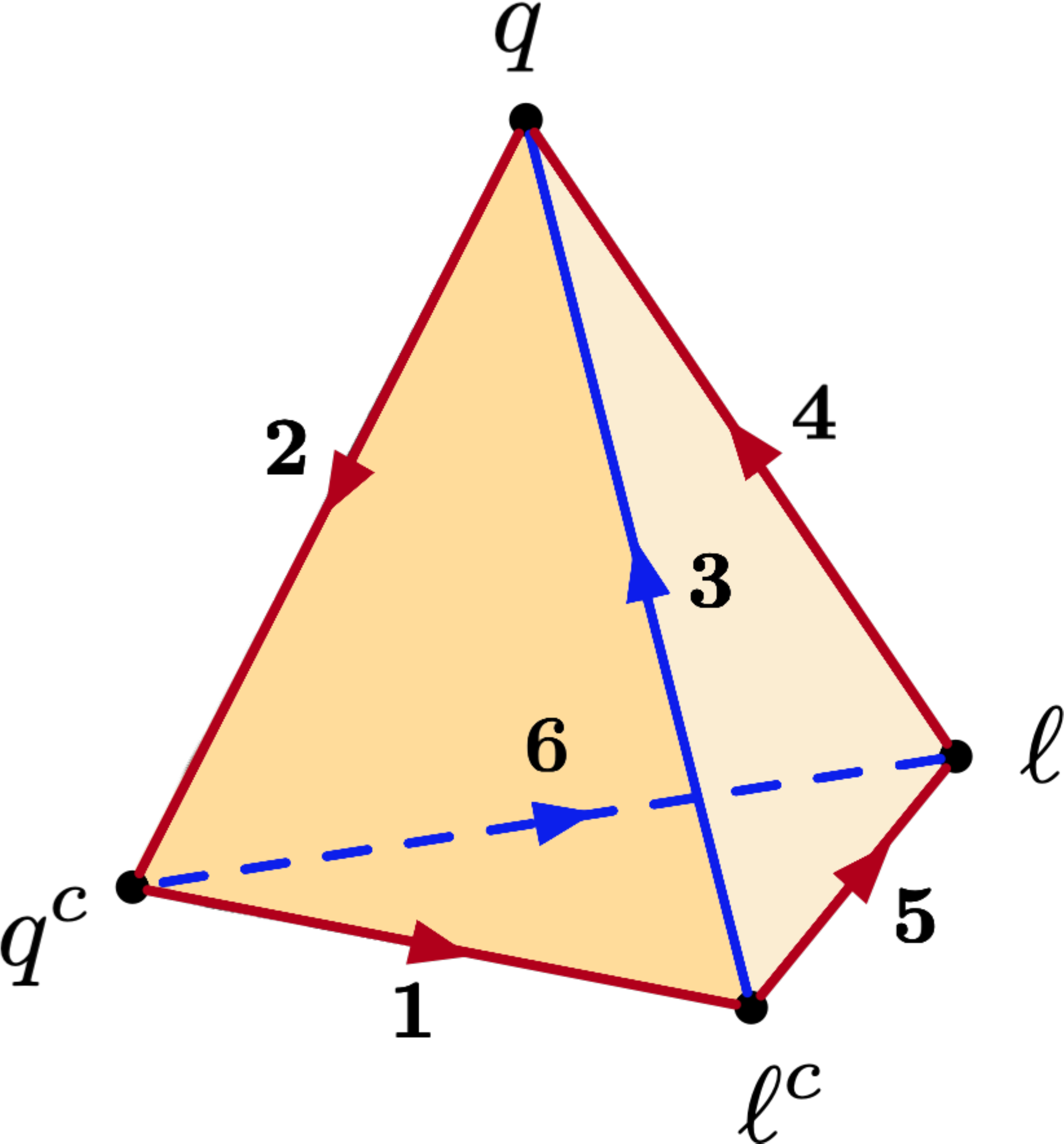}
	\caption{The Tetrahedron model
	(quartification phase): 
	The focus now is on the 4-cycle $2451$
	(red arrows) which describes a single family
	$SU(3)^4$ fermionic representation.
	It accounts for the revival of a full 
	$q,q^c\leftrightarrow \ell,\ell^c$ quark/lepton
	correspondence, as expressed by the four
	variants shown in Fig.(\ref{4variants}).
	The blue arrows account for the threefold
	family structure under horizontal
	$SU(3)_I \otimes SU(3)_{II}$.}
	\label{4cycle} 
\end{figure}

\section{
Quark/Lepton correspondence resurrection}

$B-L$ does not serve in our model as the forth color.
The reason is obvious: There is no Pati-Salam
$SU(4)$ to the rescue.
Quark/lepton correspondence is absent
from the trinification model as well. 
Unlike $q$ and $q^c$ which are assigned to
two different vertices, see Fig.(\ref{tri}), $\ell$
and $\ell^c$ share a common third vertex.
It is only in Ma's $SU(3)^4$ model that
quark/lepton correspondence has been revived,
at least in the sense that $\ell$ and $\ell^c$
split vertices, see Fig.(\ref{4cycle}).
But there is more to it.

The H-sector is primarily in charge of the
threefold family replication.
Denoting the quartification horizontal group
by $SU(3)_I \otimes SU(3)_{II}$, and choosing
for definiteness one of the four variants depicted
in Fig.(\ref{4variants}), we face the horizontal
assignments
\begin{equation}
	q \sim (1,3) ~, ~q^c\sim (3^\ast,1) ~,~
	\ell\sim (3,1) ~, ~ \ell^c\sim (1,3^\ast) ~.
\end{equation}
It is the pair $\{q^c,\ell\}$ which cancels the
$SU(3)_I$ anomalies, while $\{q,\ell^c\}$
takes care of the $SU(3)_{II}$ anomalies.
Such an unprecedented horizontal pairing,
which seems as a unified generalization of
Foot-Lew quark/lepton
symmetry \cite{FL}, is fully dictated by the
group theoretical structure of our tetrahedron
model, and gives a novel perspective to
the inter relations between quarks and leptons.

Another attractive aspect of quartification
has to do with the Higgs sector.
The standard model has already taught us that
quarks and leptons alike should acquire their
masses via Yukawa couplings with the one and
the same Higgs scalar.
One complex doublet suffices to govern all
masses (and may even be the only source of
spontaneous symmetry breaking).
Truly, this feature is forcefully shared by the
original $SU(3)^3$ trinification model as well.
Starting from the fermionic representation
Eq.(\ref{333}), the mass generating job is carried
out by the scalar $\phi$ which exhibits the fermion
bilinear quantum numbers
\begin{equation}
	 \phi \sim qq^c 
	 \sim\ell \ell^c\sim (1,3^\ast,3) ~. 
\end{equation}
This is strikingly not the case for the tetrahedron
model in the bi-trinification phase, for which
\begin{equation}
\begin{array}{rcc}
	& qq^c \sim (1,3^\ast,3 || 1,3,3^\ast) &    \\
	& \ell \ell^c \sim (1,3^\ast,3 || 3^\ast,1,1)  &   
\end{array}
\end{equation}
The more so, once $\chi$ has entered the game,
we cannot avoid having $(\ell+ \ell^c) \chi \sim
(1,3,3^\ast ||1,3^\ast,3)$ as well.
This seems to suggest that  $\ell$ and $\ell^c$
better not share a common tetrahedron vertex.
Indeed, one is back on safe (and quite attractive)
grounds provided the tetrahedron model in adopted
in its the quartification phase.

Choosing for the sake of definiteness a
particular horizontal variant, say
\begin{equation}
	\begin{array}{|c ||c|c|c|c||c|c|}
	\hline q & 3 & 3^\ast & 1 & 1 & 1 & 3\\
	\hline q^c & 3^\ast & 1 & 3 & 1 & 3^\ast & 1\\
	\hline \ell & 1 & 3 & 1 & 3^\ast & 3 & 1\\
	\hline \ell^c & 1 & 1 & 3^\ast & 3 & 1 & 3^\ast\\
	\hline \end{array} ~,
\end{equation}
quartification naturally offers the  conjugate
fermion bilinear quantum numbers
\begin{equation}
\begin{array}{rcc}
	& qq^c \sim \phi 
	\sim (1,3^\ast,3,1 || 3^\ast,3) &    \\
	& \ell\ell^c \sim \phi^\ast
	\sim (1,3,3^\ast,1 || 3,3^\ast)  &   
\end{array}
\end{equation}
In turn, as dictated by their revived correspondence,
quarks and leptons do share now a common Yukawa
coupling origin.
The associated VEV pattern needs not be simple, to
say the least, reflecting the double spinorial structure
(expressed by $i,j=1,2,3$) of
$\langle \phi\rangle =v_{L,R}^{ij}$ under $SU(3)_{I,II}$
of each of the two (LR symmetric) Weinberg-Salam
doublets.
To add a complication to the list, notice that generically
the two VEV matrices $v_L^{ij}$ and $v_R^{ij}$ cannot
be diagonalized simultaneously.
This is apparently the reason underlying the fermion
mixing phenomenon, but unfortunately not even
a tentative mixing formula can be derived at this
preliminary stage.
It is highly suggestive that $SU(3)_I \otimes SU(3)_{II}$
gets broken at a certain stage down to $SU(3)_{I+II}$,
under which $\phi$ transforms as a singlet,
with all horizontal fine details ripped off.

\section{Epilogue}

In this paper, unfortunately and admittedly, we
have offered no detailed insight into the structure
of the Fermi mass matrix.
While this is undoubtedly a drawback, it should be
gracefully criticized, recalling that after forty years
of flavor puzzle frustration, and after so many imaginative
yet unripe theoretical ideas per year, no one can so
far claim even the slightest of victories.
The major clues, so we would like to believe, rowing
against the conventional stream, are already with us.
Namely, the challenging total number three of fermion
families, and the striking failure of all grand unification
trails (string theory and extra dimensions included)
to account for it. 
The naive yet powerful idea that all threes which
accompany flavor physics must have the one and
the same gauge theoretical origin then simply paves
the way for Vertical-Horizontal symmetric trinification.
We can only hope that the resulting hereby presented
tetrahedron of flavors, stemming from trinification
but eventually focused on its quartification (rather
than bi-trinification) phase, would shed some light on the
quark/lepton correspondence and elegantly account
for the threefold family replication, at least at the group
theoretical level.

\acknowledgments
{}

\end{document}